\documentstyle[preprint,aps]{revtex}

\begin{document}
\title{Anomalous peak in the superconducting condensate density of cuprate high T$%
_{c}$ superconductors at a unique critical doping state}
\author{C. Bernhard$^{1)}$, J.L. Tallon$^{2)}$, Th. Blasius$^{3)}$, A. Golnik$^{4)}$%
, and Ch. Niedermayer$^{3)}$}
\author{1) Max-Planck-Institut f\"{u}r Festk\"{o}rperforschung Heisenbergstrasse 1,}
\address{D-70569 Stuttgart, Germany\\
2) Industrial Research Ltd., P.O. Box 31310, Lower Hutt, New Zealand \\
3) Universit\"{a}t Konstanz, Fakult\"{a}t f\={u}r Physik, D-78434\\
Konstanz, Germany\\
4.) Institute of Experimental Physics, Warsaw University, Ho\.{z}a 69,\\
00-681 Warsaw, Poland.}
\date{\today}
\maketitle

\begin{abstract}
The doping dependence of the superconducting condensate density, n$_{s}^{o}$%
, has been studied by muon-spin-rotation for Y$_{0.8}$Ca$_{0.2}$Ba$_{2}($Cu$%
_{1-z}$Zn$_{z}$)$_{3}$O$_{7-\delta }$ and Tl$_{0.5-y}$Pb$_{0.5+y}$Sr$_{2}$Ca$%
_{1-x}$Y$_{x}$Cu$_{2}$O$_{7}$. We find that n$_{s}^{o}$ exhibits a
pronounced peak at a unique doping state in the slightly overdoped regime.
Its position coincides with the critical doping state where the normal state
pseudogap first appears depleting the electronic density of states. A
surprising correlation between n$_{s}^{o}$ and the condensation energy U$%
_{o} $ is observed which suggests unconventional behavior even in the
overdoped region.
\end{abstract}

\bigskip \allowbreak

\noindent The superconducting condensate density, n$_{s}$, is proportional
to the squared amplitude of the superconducting (SC) order parameter (OP),
i.e. of the macroscopic wave function which describes the SC charge
carriers. It is thus a fundamental parameter whose variation as a function
of temperature (T) and of carrier doping provides important information
about the SC state. From early on in the investigation of the cuprate high T$%
_{c}$ superconductors (HTCS), the absolute value of n$_{s}$ has been studied
by transverse-field muon-spin-rotation (TF-$\mu $SR) measurements on
polycrystalline samples. Using this technique a linear relationship between
the low-T value, n$_{s}^{o},$ and the critical temperature, T$_{c}$, has
been established in underdoped HTCS (so-called `Uemura-line') \cite{Uemura1}%
. This finding has stimulated the development of the so-called precursor
pairing model whose basic idea is that the low value of n$_{s}^{o}$ allows
for large thermal phase fluctuations of the OP \cite{Emery1}. These phase
fluctuations can suppress the formation of a macroscopically coherent SC
state over a significant T-interval below the mean-field transition
temperature, T$_{c}^{macr}$%
\mbox{$<$}%
\mbox{$<$}%
T$_{c}^{mf}$. This precursor pairing model could also explain the so-called
pseudogap effect, which manifests itself as a partial suppression of the
low-energy charge and spin excitations in the normal state of underdoped
HTCS. In this model the pseudogap state is thought of as the macroscopically
incoherent SC state within the range T$_{c}^{macr}$%
\mbox{$<$}%
T%
\mbox{$<$}%
T$_{c}^{mf}$. \cite{Emery1} Contrasting the precursor pairing model a number
of alternative models have been proposed which associate the pseudogap state
with electronic correlations which compete with SC \cite{Loram2,diCastro1}.
In particular, it has been shown that the combined suppression of T$_{c}$
and n$_{s}^{o}$ in underdoped HTSC can be equally well explained in terms of
the depletion of the density of states near the Fermi-level, N($\epsilon
_{F}\pm \Delta $), brought about by competing correlations which rapidly
grow in strength on the underdoped side \cite{Loram1}. The question as to
the origin of the normal state pseudogap is presently vigorously debated and
is considered as an important key to resolve the mystery of HTCS.

In this paper we report extensive TF-$\mu $SR studies on the evolution of n$%
_{s}^{o}/m_{ab}^{\ast }$ as a function of hole doping, p, for series of
polycrystalline Y$_{0.8}$Ca$_{0.2}$Ba$_{2}$(Cu$_{1-z}$Zn$_{z}$)$_{3}$O$%
_{7-\delta }$ (Y,Ca-123) and Tl$_{0.5-y}$Pb$_{0.5+y}$Sr$_{2}$Ca$_{1-x}$Y$%
_{x} $Cu$_{2}$O$_{7}$ (Tl-1212) samples. Our new data complement previous
less detailed studies \cite{Niedermayer1,Bernhard1,Tallon1,Bernhard2} and
demonstrate that n$_{s}^{o}$ exhibits a pronounced peak at a unique doping
state in the slightly overdoped region. Most remarkably, the location of the
maximum of n$_{s}^{o}$ coincides with the critical doping state where
previously a rapid suppression of N($\epsilon _{F}\pm \Delta $) has been
observed signaling the onset of competing pseudogap correlations \cite
{Loram2,Loram1}. We argue that the sudden and pronounced decrease of n$%
_{s}^{o}$ below critical doping cannot readily be explained in terms of the
precursor pairing model for which a smooth crossover would be expected. In
addition, the strong decrease of n$_{s}^{o}$ on the overdoped side indicates
unconventional behavior even in the absence of the pseudogap correlations.

Series of under- to overdoped polycrystalline samples of Y$_{0.8}$Ca$_{0.2}$%
Ba$_{2}$(Cu$_{1-z}$Zn$_{z}$)$_{3}$O$_{7-\delta }$ with 0.04$\leq \delta \leq 
$0.98 and z=0, 0.02 and 0.04 and Tl$_{0.5-y}$Pb$_{0.5+y}$Sr$_{2}$Ca$_{1-x}$Y$%
_{x}$Cu$_{2}$O$_{7}$ with y$\leq $0.15 and x$\leq $0.4 have been prepared
following previously described procedures \cite{Bernhard1,Tallon1,Presland1}%
. The T$_{c}$ values have been determined by resistivity and DC
susceptibility measurements. The hole doping of the CuO$_{2}$ planes, p, has
been deduced from measurements of the room-temperature thermo-electric power
(RT-TEP) \cite{Obertelli1}. Alternatively, p has been estimated from the
ratio of T$_{c}$/T$_{c,max}$ (knowing, e.g., from the RT-TEP, whether the
sample is under- or overdoped) by assuming the approximate parabolic
p-dependence in which p=0.16$\pm \sqrt{(1-T_{c}/T_{c,\max })/82.6}$\cite
{Presland1,Tallon2}. Good agreement has been obtained between both estimates.

The TF-$\mu $SR experiments at an external field of 3 kOe have been
performed at the $\pi $M3 beamline of the muon facility of the
Paul-Scherrer-Institut (PSI) in Villigen/Switzerland. A detailed description
of the TF-$\mu $SR technique and its use in determining n$_{s}$ for
polycrystalline HTSC samples is given in reference \cite{Puempin1}. A
Gaussian relaxation function has been used to fit the measured time spectra.
From the obtained Gaussian depolarisation rate, $\sigma ,$ we deduced the
magnetic penetration depth, $\lambda _{ab}$, and the ratio of n$_{s}$ to the
effective carrier mass m$_{ab}^{\ast }$ using the established relationship 
\cite{Bernhard1,Puempin1}: $\sigma \;[\mu s^{-1}]=7.086\cdot 10^{4}\cdot
\lambda _{ab}^{-2}\;[nm]=2.51\cdot 10^{-21}\cdot m_{e}\cdot
n_{s}/m_{ab}^{\ast }\;[cm^{-3}].$

Figure 1(a) displays the evolution of the low-T value of the depolarisation
rate, $\sigma _{o}\sim $n$_{s}^{o}/m_{ab}^{\ast },$ as a function of the
hole doping per CuO$_{2}$ plane, p, for the series of under- to overdoped Y$%
_{0.8}$Ca$_{0.2}$Ba$_{2}$(Cu$_{1-z}$Zn$_{z}$)$_{3}$O$_{7-\delta }$ with z=0,
0.02 and 0.04. Figure 1(b) shows the evolution of T$_{c}$ with p. It is
evident from Fig. 1 that n$_{s}^{o}$ exhibits a pronounced peak in the
slightly overdoped regime. For all three series it occurs at a similar
doping state of p $\approx $ 0.19 (in the following we call it `critical
doping' p$_{crit}$). At optimum doping of p$_{opt}\approx $0.16, where the
highest T$_{c}$ value of T$_{c,max}$=85.5 K is observed for the Zn-free
series, n$_{s}^{o}$ is already reduced by 25-30 \% as compared to critical
doping. The difference between optimum doping (highest T$_{c}$ value) and
critical doping (highest n$_{s}^{o}$ value) is largest for the pure series,
but it is reduced for the Zn-substituted series since the optimum doping
concentration increases upon Zn-substitution. At very high Zn content, SC
survives only in the vicinity of the critical doping state, i.e. p$%
_{opt}\rightarrow $p$_{crit}$. This effect has been previously explained in
terms of the suppression of N($\varepsilon _{F}\pm \Delta $) due to the
opening of the pseudogap below critical doping which, for impurity
scattering in the unitarity limit, enhances the suppression of T$_{c}$ and
of n$_{s}^{o}$ \cite{Tallon4}. The other remarkable feature is the plateau
in n$_{s}^{o}$ versus p centered around 1/8 doping. The p-dependence of n$%
_{s}^{o}$ therefore is characterized by two marked features, a plateau
around 1/8 doping (likely associated with the formation of static stripes)
and the peak near critical doping (due to some kind of yet unknown
electronic or magnetic correlations which compete with SC). Note that for
these Y,Ca-123 samples the contribution of the CuO chains to n$_{s}^{o}$
should be much weaker than in YBa$_{2}$Cu$_{4}$O$_{8}$ (Y-124) or in fully
oxygenated Y-123 since the CuO chains are significantly deoxygenated except
for the strongly overdoped regime and even there the Ca-substitution leads
to chain disorder by occupation of the O(5) off-chain position \cite
{Bernhard1,Tallon1,Tallon2}.

In order to confirm that the peak of n$_{s}^{o}$ at p$_{crit}\approx $0.19
is a common feature of the hole-doped HTCS we have also investigated a
series of under- to overdoped Tl-1212 polycrystalline samples (T$_{c,max}$%
=107 K). Figure 2 shows the evolution of $\sigma _{o}\sim $n$%
_{s}^{o}/m_{ab}^{\ast }$ as a function of p. It is evident that $\sigma _{o}$
follows a similar p-dependence as in Y,Ca-123 and, in particular, that it
also exhibits a peak at p$_{crit}\approx $0.19.

It is remarkable that the same critical doping state has been previously
identified based on specific heat, susceptibility and NMR data as the point
where the NS pseudogap first appears and starts to deplete N($\epsilon
_{F}\pm \Delta $) \cite{Loram2,Loram1,Tallon5}. In Fig. 3 the solid line
marked by the crosses shows the p-dependence of the ratio of the electronic
entropy divided by the temperature, (S/T)$_{T_{c}}$=($\stackrel{T_{c}}{\int }%
\gamma \left( T^{\prime }\right) dT^{\prime })/T_{c},$ (normalized to the
value at critical doping) as obtained by Loram et al. from specific heat
data for a similar Y,Ca-123 series \cite{Loram1,Loram3}. (S/T)$_{T_{c}}$ is
the average of $\gamma $ between T=0 and T$_{c}$ and is a measure of the
average density of states within an energy window of $\sim 2-3$\ k$_{B}$T$%
_{c}$ and thus proportional to N($\epsilon _{F}\pm \Delta $) just above T$%
_{c}$. It is almost constant on the strongly overdoped side while it
exhibits a steady decrease below critical doping due to the opening of the
pseudogap. A very similar p-dependence has been obtained from optical
measurements for the plasma frequency, $\omega _{pl}^{n}$, of the normal
carriers \cite{Puchkov1}. Shown by the dotted line and marked by the stars
is the condensation energy U$_{o}$=$\int^{T_{c}}(S_{n}-S_{s})dT^{\prime }$
normalized to its value at critical doping \cite{Loram1,Loram3}. Finally,
the normalised condensate density n$_{s}^{o}$/n$_{s,max}^{o}$ of the pure
Y,Ca-123 and the Tl-1212 series reported here is shown by the open circles
and solid triangles, respectively. It is evident from Fig. 3 that n$_{s}^{o}$
and U$_{o}$ follow very similar doping dependencies. Both of them exhibit a
pronounced peak around critical doping, they decrease rather steeply on the
underdoped as well as on the overdoped sides. Such a correlation is expected
on the underdoped side due to the decreasing density of the normal carriers.
In fact, (S/T)$_{T_{c}}$ and n$_{s}^{o}$ can be seen to match very well
below p$_{crit}$. Also, the decrease of U$_{o}$ on the overdoped side can be
understood within a BCS-model due to the decreasing size of T$_{c}$ and the
concomitant decrease of the size of the SC energy gap (such as observed by
spectroscopic techniques \cite{White1,Bernhard3}). In clear contrast, the
dramatic decrease of n$_{s}^{o}/m_{ab}^{\ast }$ on the overdoped side is
completely unexpected since the electronic density of states $\sim $(S/T)$%
_{T_{c}}$ or likewise the plasma frequency of the normal carriers $\omega
_{pl}^{n}\sim n_{n}/m_{ab}^{\ast }$ \cite{Puchkov1} remain almost constant
on the overdoped side. We return to this important point later and focus
first on the observed behavior on the underdoped side. Below critical doping
it is evident that all three quantities which are displayed in Fig. 3 follow
a common doping dependence, i.e. below p$_{crit}$ they suddenly start to
decrease. It has been pointed out earlier that this circumstance of a steady
reduction of N($\varepsilon _{F}\pm \Delta $) for p%
\mbox{$<$}%
p$_{crit}$ accompanied by a sharp reduction in the condensation energy U$%
_{o} $ and the condensate density n$_{s}^{o}$ is precisely what is expected
with the onset of an electronic correlation competing with SC \cite
{Loram2,Loram1}. Note that n$_{s}^{o}$ characterizes the ground state
property of a SC. Its sudden change at p$_{crit}$ is thus indicative of a
drastic alteration of the SC ground-state which is not expected for the
precursor pairing model where instead a smooth crossover should occur \cite
{Emery1}.

The argument against precursor pairing can be further substantiated by
examples which show that T$_{c}$ of underdoped samples is not uniquely
determined by n$_{s}^{o}$. YBa$_{2}$Cu$_{4}$O$_{8}$ (Y-124) has, besides the
superconducting CuO$_{2}$ bilayers, metallic CuO chains that become
superconducting most likely due to proximity coupling \cite
{Bernhard1,Tallon1}. Y-124 is underdoped with T$_{c}$=80 K and exhibits
clear signatures of the pseudogap. The metallic CuO double chains, however,
lead to a significant enhancement of the condensate density (with $\sigma
_{o}$=3.3 $\mu $s$_{s}^{-1})$ as compared to similarly underdoped Y-123 or
Y,Ca-123 with T$_{c}\approx $80 K and $\sigma _{o}\sim $2.2-2.3 $\mu $s$%
^{-1} $ \cite{Bernhard1,Tallon1}.\ If T$_{c}$ of underdoped samples was
indeed uniquely determined by n$_{s}^{o}$, then T$_{c}$ should exceed 100 K
in Y-124$.$ Another counter example is underdoped Zn-substituted Y-123. It
has been shown that upon Zn-substitution n$_{s}^{o}$ is even more rapidly
suppressed than T$_{c}$. This behavior has been explained in terms of the
d-wave symmetry of the superconducting OP and elastic scattering in the
unitarity limit on Zn-impurities \cite{Bernhard2}. In a plot of T$_{c}$
versus n$_{s}^{o}$ such compounds thus lie far to the left of the `Uemura
line'. For Y$_{0.8}$Ca$_{0.2}$Ba$_{2}$Cu$_{2.94}$Zn$_{0.06}$O$_{6.45}$ with T%
$_{c}=$31 K we obtained $\sigma _{o}\approx $0.2 $\mu $s$_{s}^{-1}$ which
according to the Uemura-relation should result in T$\leq $10 K, i.e. one
third the observed value \cite{Bernhard2}. Finally, there is the surprising
result that n$_{s}^{o}$ decreases very strongly also on the overdoped side 
\cite{Niedermayer1}. Despite the small n$_{s}^{o}$ values no pseudogap
effect is observed for strongly overdoped samples. These examples imply that
it is not primarily the small value of n$_{s}^{o}$ which is responsible for
the low T$_{c}$ values of underdoped HTCS or which causes the pseudogap
effect. This conclusion is in clear contrast to the precursor pairing model.

Finally, we focus on the TF-$\mu $SR data on the strongly overdoped side
past critical doping. Our measurements confirm previous reports that n$%
_{s}^{o}/m_{ab}^{\ast }$ is dramatically reduced on overdoping. This very
surprising result was first obtained by TF-$\mu $SR on Tl$_{2}$Ba$_{2}$CuO$%
_{6+\delta }$ (Tl-2201) \cite{Niedermayer1} and later on (Yb,Ca)-123 and
(Y,Ca)-123 \cite{Bernhard1,Tallon1}. Subsequently, it has been confirmed by
other experimental techniques \cite{Locquet1}. It was previously pointed out
that the strong suppression of n$_{s}^{o}/m_{ab}^{\ast }$ for heavily
overdoped samples cannot be understood within a BCS model, unless one
assumes that pairbreaking correlations become increasingly important \cite
{Niedermayer1,Bernhard2}. This earlier proposal of strong pair breaking on
the overdoped side, however, is not supported by recent specific heat \cite
{Loram1} or $^{89}$Y- and $^{17}$O-NMR\ data \cite{Williams1} which give no
clear indication for a growing density of unpaired carriers within the SC
gap. These data also do not support the scenario that overdoped materials
are inhomogeneous with only a small SC fraction \cite{Uemura2}. As was
mentioned above, the dramatic decrease of n$_{s}^{o}/m_{ab}^{\ast }$ in the
overdoped region is completely unexpected. The BCS-model predicts that n$%
_{s}^{o}$/m$_{ab}^{\ast }$ should follow the same p-dependence as the normal
state plasma frequency, $\omega _{pl}^{n}\sim n_{n}/m_{ab}^{\ast },$ deduced
from optical experiments \cite{Puchkov1} or (S/T)$_{T_{c}}$ obtained from
the specific heat \cite{Loram2,Loram3} which both remain almost constant on
the overdoped side. In this context we would like to emphasize the
surprising similarity between the p-dependences of n$_{s}^{o}$ as deduced
from TF-$\mu $SR \ and the condensation energy U$_{o}$ obtained from the
specific heat measurements \cite{Loram1,Loram3}. Figure 3 implies that $%
n_{s}^{o}$ and U$_{o}$ are correlated over the entire doping range. Such a
relationship cannot easily be understood within BCS theory where the
condensation energy U$_{o}$ is determined by the change in potential energy
due to the attractive pairing interaction times the density of state N($%
\varepsilon _{F}\pm \Delta $), while n$_{o}^{s}$ should be determined by the
kinetic energy of the carriers times N($\varepsilon _{F}\pm \Delta $). This
apparent inconsistency of the experimental data on the overdoped side with
the prediction of the BCS-model is especially important in the light of the
indication that competing electronic correlations are at work below critical
doping. This should mean that the intrinsic properties of the SC state are
best seen for overdoped materials. Yet it is this very region which cannot
be described by the BCS-model and therefore suggest an unconventional SC
state. As one example of an unconventional model which explains the unusual
correlation between U$_{o}$ and n$_{s}^{o}$ we only mention here the
so-called spin-charge separation model. Superconductivity can only occur
here if both holons and spinons condense resulting in a total condensate
density of 1/n$_{s}$=1/n$_{s}^{holon}$+1/n$_{s}^{spinon}$ \cite{Lee1}. On
the overdoped side the diminishing spinon density would therefore lead to
the dramatic reduction of n$_{s}^{o}$. As a final remark we note that a
similar TF-$\mu $SR study has recently been performed on the p-dependence of
n$_{s}^{o}$ in La$_{2-x}$Sr$_{x}$CuO$_{4}$ (La-214) \cite{Panagopoulos1}.
For La-214 the maximum in n$_{s}^{o}$ versus p is significantly broader than
the one reported here for Y,Ca-123 and Tl-1212. The reason for this probably
lies in the fact that N($\varepsilon _{F}\mp \Delta $) for La-214 goes on
increasing with overdoping \cite{Loram3}. What is important is that while n$%
_{s}^{o}$/(S/T)$_{T_{c}}$ for La-214 remains constant below p$_{crit}$, this
ratio decreases sharply for p%
\mbox{$>$}%
p$_{crit}$\ just as shown in Fig. 3 for Y,Ca-123 and Tl-1212. Moreover, the
peak in U$_{o}$ for La-214 is also significantly broadened as compared with
the one in Y,Ca-123 \cite{Loram3}. This finding seems to support our
suggestion that n$_{s}^{o}$ and U$_{o}$ may be correlated in the HTSC.

In summary, we have presented experimental evidence based on TF-$\mu $SR
measurements that a unique critical doping state exists in the slightly
overdoped regime where the superconducting condensate density, n$_{s}^{o}$,
exhibits a pronounced maximum. The observed sudden change of n$_{s}^{o}$ is
not expected within the precursor pairing model \cite{Emery1}. On the
underdoped side the decrease of n$_{s}^{o}$ coincides with the decrease of
the normal density of states caused by the pseudogap correlations. This
effect sets in abruptly at critical doping \cite{Loram1,Tallon5,Loram3}. In
contrast, the rapid decrease of n$_{s}^{o}$ on the strongly overdoped side
cannot be easily understood within a conventional BCS model since the normal
state carrier density remains almost constant here. Even more surprisingly, n%
$_{s}^{o}$ seems to scale with the condensation energy U$_{o}$ as obtained
from specific heat measurements \cite{Loram1,Loram3}. This correlation
suggests unconventional behavior even in the overdoped region.

We acknowledge A. Amato and D. Herlach (PSI) for technical support. C.B. and
J.L.T thank J.W. Loram for fruitful discussions and for providing the
specific heat data. C.B. appreciates financial support of the Marsden Fund
of New Zealand during his stay at IRL. J.L.T. thanks the Royal Society of
New Zealand for financial support under a James Cook Fellowship.

{\bf Figure Captions}

\bigskip

Figure 1: (a) Evolution of the low-T depolarisation rate $\sigma _{o}\sim $n$%
_{s}^{o}/m_{ab}^{\ast }$ as a function of hole doping, p, for series of
under- to overdoped Y$_{0.8}$Ca$_{0.2}$Ba$_{2}$Cu$_{3-z}$Zn$_{z}$O$%
_{7-\delta }$ with z=0 (open circles), z=0.02 (solid squares) and z=0.04
(stars). The crossed circle shows $\sigma _{o}$ for strongly overdoped Lu$%
_{0.7}$Ca$_{0.3}$Ba$_{2}$Cu$_{3}$O$_{6.95}$. Critical (optimum) doping is
marked by the solid (dotted) line. (b) Doping dependence of the critical
temperature T$_{c}$ shown by the same symbols.

\bigskip

Figure 2: Doping dependence of the low-T depolarisation rate $\sigma _{o}$
for under- to overdoped Tl$_{1-y}$Pb$_{y}$Sr$_{2}$Ca$_{1-x}$Y$_{x}$Cu$_{2}$O$%
_{7}$. The solid (dotted) line marks critical (optimum) doping.

\bigskip

Figure 3: Doping dependence of the normalized values of (S/T)$_{T_{c}}$
(solid line and crosses) and of U$_{o}$ (solid line and stars) of Y,Ca-123
as deduced from specific heat \cite{Loram1,Loram3} and of n$_{s}^{o}$/m$%
_{ab}^{\ast }$ for Y,Ca-123 (open circles) and Tl-1212 (solid triangles)
deduced from TF-$\mu $SR.

\end{document}